\newcommand{\ergcms}{{\rm ergs}\ {\rm cm}^{-2}\ {\rm s}^{-1}}

\documentclass{pasj00}
\SetRunningHead{M. Nobukawa et al.}{Suzaku Observation of the Sgr~B1 Region}
\Received{2007/05/04}
\Accepted{2007/08/27}
\Published{published}
\begin{document}
\title{Suzaku Spectroscopy of an X-Ray Reflection Nebula and 
a New Supernova Remnant Candidate in the Sgr B1 Region}
\author{Masayoshi \textsc{Nobukawa}, Takeshi Go \textsc{Tsuru}, 
Yojiro \textsc{Takikawa}, Yoshiaki \textsc{Hyodo}, Tatsuya \textsc{Inui}, \\
Hiroshi \textsc{Nakajima}, Hironori \textsc{Matsumoto} and Katsuji \textsc{Koyama}, }
\affil{Department of Physics, Graduate School of Science, Kyoto University, 
Sakyo-ku, Kyoto 606-8502}
\email{nobukawa@cr.scphys.kyoto-u.ac.jp}
\author{Hiroshi \textsc{Murakami}}
\affil{PLAIN center, ISAS/JAXA, 3-1-1 Yoshinodai, Sagamihara, Kanagawa 229-8510}
\and
\author{Shigeo \textsc{Yamauchi}}
\affil{Faculty of Humanities and Social Sciences, Iwate University, 3-18-34 Ueda, 
Morioka, Iwate 020-8550}
\maketitle

%%%%%%%%%%%%%%%%%%%%%%%%%%%%%%%%%%%%%%%%%%%%%%%%
%%%%%%%%%%%%%%%%%%%%%%%%%%%%%%%%%%%%%%%%%%%%%%%%

\begin{abstract}

We made a 100~ks observation of the Sagittarius (Sgr) B1 region at 
$(l,\ b)=(\timeform{0\circ.5},\ \timeform{-0\circ.1})$ near to the Galactic 
center (GC) with the Suzaku/XIS. 
Emission lines of S\emissiontype{XV}, Fe\emissiontype{I} 
, Fe\emissiontype{XXV}, and Fe\emissiontype{XXVI} 
were clearly detected in the spectrum. We found that the Fe\emissiontype{XXV} 
and Fe\emissiontype{XXVI} line emissions smoothly distribute over the Sgr~B1 
and B2 regions connecting from the GC. This result 
suggests that the GC hot plasma extends at least up to the Sgr~B 
region with a constant temperature. There are two diffuse X-ray sources 
in the observed region. One of the two (G0.42$-$0.04) 
is newly discovered, and exhibits a strong S\emissiontype{XV} K$\alpha$ emission 
line, suggesting a candidate for a supernova remnant located in the GC region. 
The other one (M0.51$-$0.10), having a prominent 
Fe\emissiontype{I} K$\alpha$ emission line and a strongly absorbed continuum, is 
likely to be an X-ray reflection nebula. There is no near source bright 
enough to irradiate M0.51$-$0.10. However, the Fe\emissiontype{I} K$\alpha$
emission can be explained if Sgr~A* was $\sim10^6$ times brighter $300$ years ago, 
the light travel time for 100 pc to M0.51$-$0.10, than it is at present. 
\end{abstract}

%%%%%%%%%%%%%%%%%%%%%%%%%%%%%%%%%%%%%%%%%%%%%%%%
%%%%%%%%%%%%%%%%%%%%%%%%%%%%%%%%%%%%%%%%%%%%%%%%

\section{Introduction}

One of the remarkable discoveries from X-ray observations of the Galactic 
center (GC) region is the Galactic center diffuse X-ray emission (GCDX) 
extending over $\sim$300 pc ($|l|<\timeform{1\circ}$), and that the GCDX has 
a prominent highly ionized iron line at 6.7 keV \citep{Koyama1989, Yamauchi1990}. 
ASCA detected Fe\emissiontype{XXV} and Fe\emissiontype{XXVI} emission lines 
from the GCDX, and its spectrum could interpreted to be emitted from a hot plasma 
of $kT \sim 10$ keV \citep{Koyama1996}. Chandra, with an extremely high spatial 
resolution of $\sim\timeform{0''.5}$, has detected many X-ray point sources 
in the GC region. The total flux of the detected point sources is not enough to 
explain all of the GCDX \citep{Muno2004GCdiffuse}. On the other hand, 
\citet{Revnivtsev2006} indicate that the X-ray emission can be completely 
explained by integration of point sources. Suzaku revealed that the origin of the 
iron lines is not the charge-exchange of cosmic-rays with molecular clouds, but 
collisional ionization of hot plasma with a temperature of $kT\sim6.5$ keV 
\citep{Koyama2007GCplasma}. Moreover, the spatial distribution is different 
from that of point sources observed by Chandra, suggesting that the main part 
of the emission is from truly diffuse plasma in $l=\timeform{-0\circ.4}-
\timeform{+0\circ.2}$ \citep{Koyama2007GCplasma}. Where the hot plasma has 
extended is not yet clear.

In addition, the origin of the huge thermal energy ($10^{53-54}$ ergs) of the 
hot diffuse plasma is a serious mystery. One hypothesis is a multiple supernova 
explosions scenario in which the dynamical timescale of $10^5$ years for the hot 
plasma requires tens of supernova explosions during a period of ten thousand 
years. Since a few supernova remnants (SNRs) have been detected so far, 
there would exist many undiscovered SNRs. 

Another remarkable discovery is diffuse emission of Fe\emissiontype{I} 
at 6.4~keV. It was firstly detected by ASCA. In particular, bright emission 
regions of the 6.4~keV line have consistent positions of the giant molecular 
clouds of Sgr~B2, Sgr~C, and the Radio Arc \citep{Koyama1996,Murakami2001CXOsgrB2, 
Murakami2001ASCAsgrC}. Their X-ray spectra exhibit a large equivalent width 
($\gtrsim1$ keV) for the Fe\emissiontype{I} emission line and deep absorption 
column ($\gtrsim 10^{23}$ cm$^{-2}$). This suggests that those molecular clouds 
were irradiated with hard X-rays from an external source and ``reflected'' them 
as the 6.4 keV X-rays (X-ray reflection Nebulae; XRNe). There is no irradiating 
source with a luminosity of $\gtrsim10^{39}$ ergs s$^{-1}$ to explain those 
6.4~keV X-rays in the GC region. \citet{Koyama1996} proposed a scenario of a past 
X-ray outburst of the supermassive black hole at Sgr~A*. Since its present 
luminosity is $2\times10^{33}$ ergs s$^{-1}$ \citep{Baganoff2003sgrA*}, it would 
have been $10^{6}$ times brighter than now 300 years ago.

On the other hand, \citet{Yusef2007} suggest that the 6.4 keV X-rays are due to 
collision of low-energy cosmic-ray electrons with this molecular gas because of a 
correlation between non-thermal radio filaments and the X-ray features. The 
scenario explains not only the Fe\emissiontype{I} K$\alpha$ emission, but also 
the cosmic-ray heating of molecular gas and diffuse TeV emission from the Galactic 
molecular clouds.

The radio complex Sgr~B1, located to the west, or to the negative Galactic 
longitude of the Sgr~B2 giant molecular cloud consists of a H\emissiontype{II} 
region surrounded by a molecular loop \citep{Sofue1990}. The expansion of the 
molecular loop is explained by past activity of the H\emissiontype{II} region 
\citep{Sofue1990}. A 6.4~keV clump was discovered in this region by 
\citet{Yusef2007}. OH and H$_2$O masers have been found, and many intermediate 
and/or low-mass stars are formed \citep{Mehringer1993}.

The X-ray Imaging Spectrometers (XIS) aboard Suzaku have characteristic 
features of a large effective area and a low/stable background. The XIS are 
instruments suitable to observe diffuse and faint objects, such as XRNe and SNRs. 
We had observed the Sgr~B1 region for about 100 ksec with Suzaku in order to 
obtain a good spectrum of the new 6.4~keV cloud detected by \citet{Yusef2007}. 
We report on results of the observation.

%%%%%%%%%%%%%%%%%%%%%%%%%%%%%%%%%%%%%%%%%%%%%%%%%%%%%%%%%%%%%%%%%%%%%%%%%%%%%%%%%%%
%%%%%%%%%%%%%%%%%%%%%%%%%%%%%%%%%%%%%%%%%%%%%%%%%%%%%%%%%%%%%%%%%%%%%%%%%%%%%%%%%%%

\section{Observation and Data Processing}
We observed the Sgr~B1 region with Suzaku on 2006 March 27--29. Though the Suzaku 
observation was performed with the Hard X-ray Detector \citep{Takahashi2007HXD, 
Kokubun2007HXD} and the XIS, we concentrate on the XIS data in this paper. The 
XIS system consists of four sets of CCD cameras (designated as XIS0, 1, 2, and 3) 
placed on the focal planes of the four X-Ray Telescopes (XRT), whose half-power 
diameters are $\sim\timeform{2\prime}$ \citep{Serlemitsos2007XRT}. XIS0, 2, and 3 
have front-illuminated (FI) CCDs, while XIS1 contains a backside-illuminated (BI) 
CCD. The energy resolution of the XIS was $\sim170$ eV at 5.9 keV  (calibration 
source of $^{55}$Fe) when this observation was performed. The details of Suzaku, 
XIS, and XRT are given in \citet{Mitsuda2007Suzaku}, \citet{Koyama2007XIS}, and 
\citet{Serlemitsos2007XRT}, respectively.

The XIS data were taken in the normal mode. The XIS pulse-height data for each 
X-ray event were converted to Pulse Invariant (PI) channels using the xispi 
software version 2006-12-26, and the calibration database version 2006-05-22. 
After removing the epoch of low-Earth elevation angles less than 5 degrees 
(ELV$<5^{\circ}$), the day Earth elevation angle less than 10 degrees 
(DYE{\_}ELV$<10^{\circ}$) and the South Atlantic Anomaly (SAA), the effective 
exposure time was about 95 ks. We performed the data reduction and the analysis 
using HEADAS software 6.1.2 and {\tt XSPEC}~11.3.2. In spectral fittings, we 
used the XIS response files released on 2006-08-01 by the Suzaku XIS team, and 
generated auxiliary files with {\tt xissimarfgen} in HEADAS software 6.1.2. 
Since the relative gains among FI sensors are well calibrated and the response 
functions are essentially the same, we co-added their data. We also applied 
non-X-ray background (NXBG) data from the night-Earth data released by 
the Suzaku XIS 
team.\footnote{http://www.astro.isas.jaxa.jp/suzaku/analysis/xis/nte/} The NXBG 
data were sorted with the geomagnetic cut-off rigidity (COR) because of the 
variation of the NXBG corresponding to the COR values. 

%%%%%%%%%%%%%%%%%%%%%%%%%%%%%%%%%%%%%%%%%%%%%%%%%%%%%%%%%%%%%%%%%%%%%%%%%
%%%%%%%%%%%%%%%%%%%%%%%%%%%%%%%%%%%%%%%%%%%%%%%%%%%%%%%%%%%%%%%%%%%%%%%%%

%%%%%%%%%%%%%%%%%%%%%%%%
%% Analysis & Results %%
%%%%%%%%%%%%%%%%%%%%%%%%

\section{Analysis and Results}
\subsection{The Overall Features}
\begin{figure}
  \begin{center}
    \FigureFile(80mm,60mm){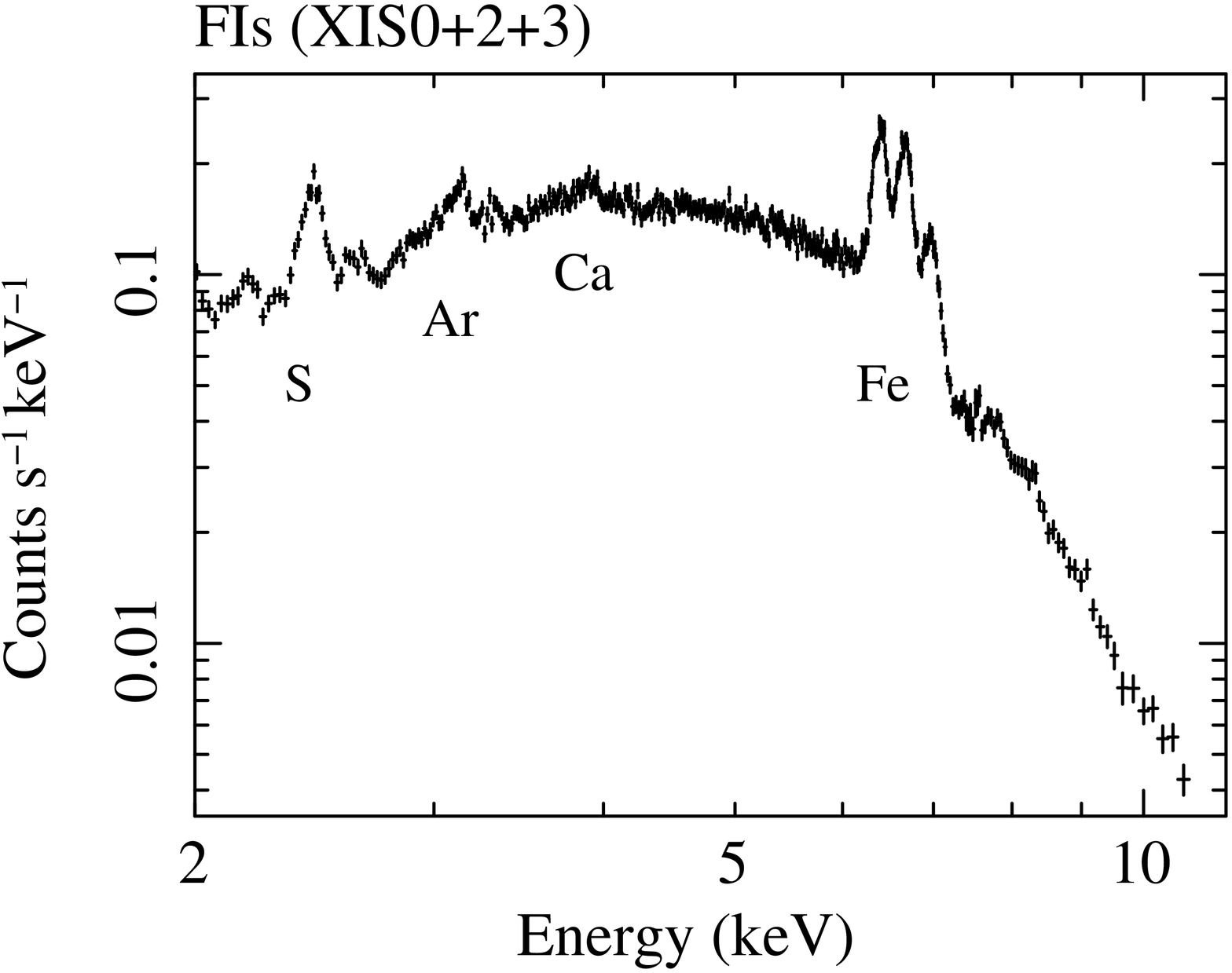}
    \FigureFile(80mm,60mm){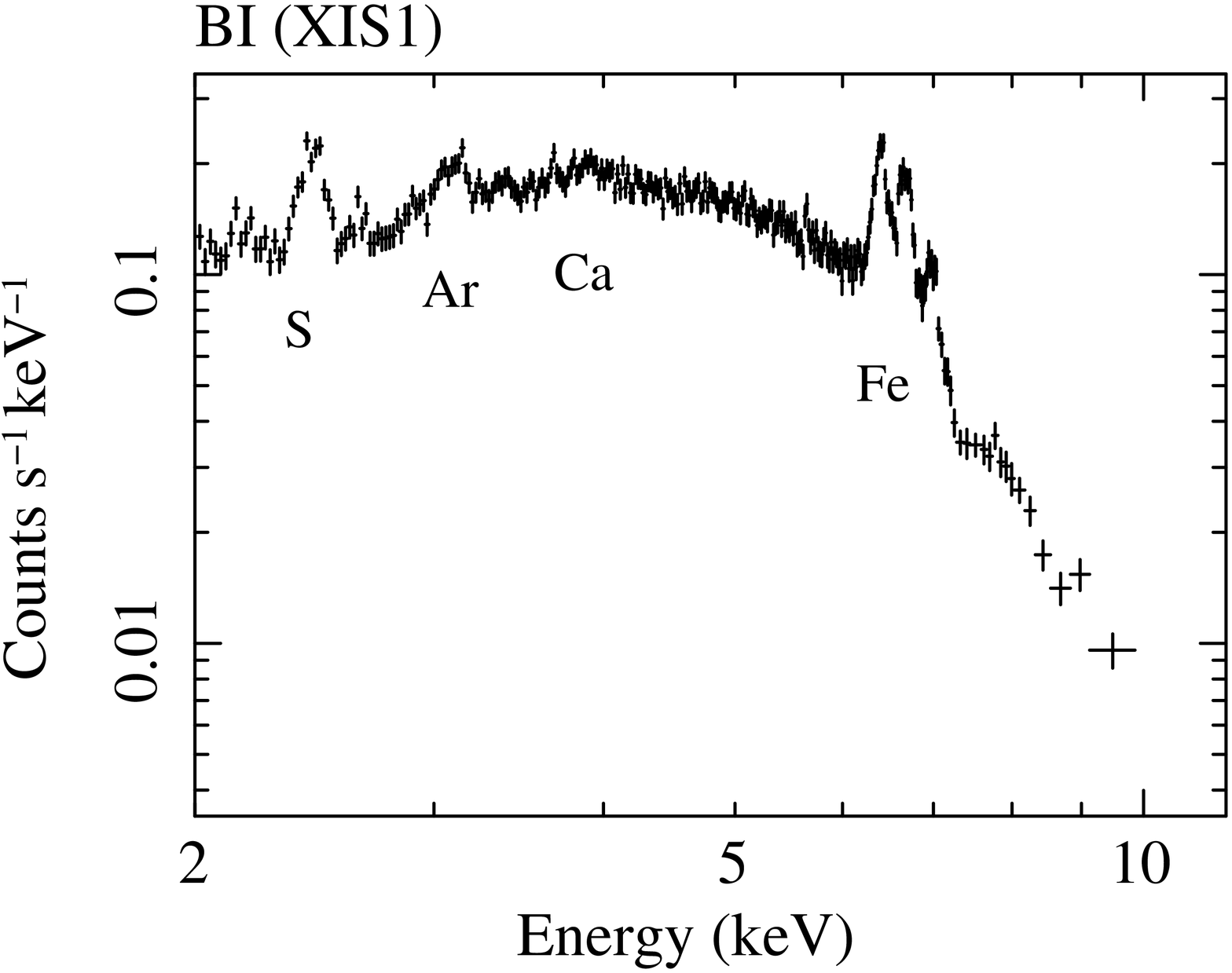}
  \end{center}
  \caption{Suzaku/XIS spectra of the Sgr~B1 region. The top and
    bottom panels show the co-added spectrum of the FIs (XIS0+2+3) and
    that of the BI (XIS1), respectively. The spectra were collected
    from the full FOV, but excluded the corners where the
    calibration sources ($^{55}$Fe) illuminate the Mn\emissiontype{I} K lines. 
    The non-X-ray backgrounds were already subtracted.}
 \label{fig:full-spectra}
\end{figure}
We show the X-ray spectra of the Sgr~B1 region in figure~\ref{fig:full-spectra}. 
They are the FI (the average of XIS0, 2, and 3;
top) and BI (XIS1; bottom) spectra extracted from the full FOV,
excluding the corners where the calibration sources of $^{55}$Fe 
illuminate. The NXBG was already subtracted from the spectra. 
They show K emission lines from He-like and/or H-like ions of S, Ar,
Ca, and Fe. In addition, the K$\alpha$ emission line at 6.4~keV and the
K-edge absorption at 7.1~keV of neutral or low ionized Fe are clearly
detected. Three Fe lines at 6.4, 6.7, and 6.9 keV are well resolved,
thanks to the excellent energy resolution of the XIS. 

\begin{figure*}[t]
  \begin{center}
    \FigureFile(160mm,50mm){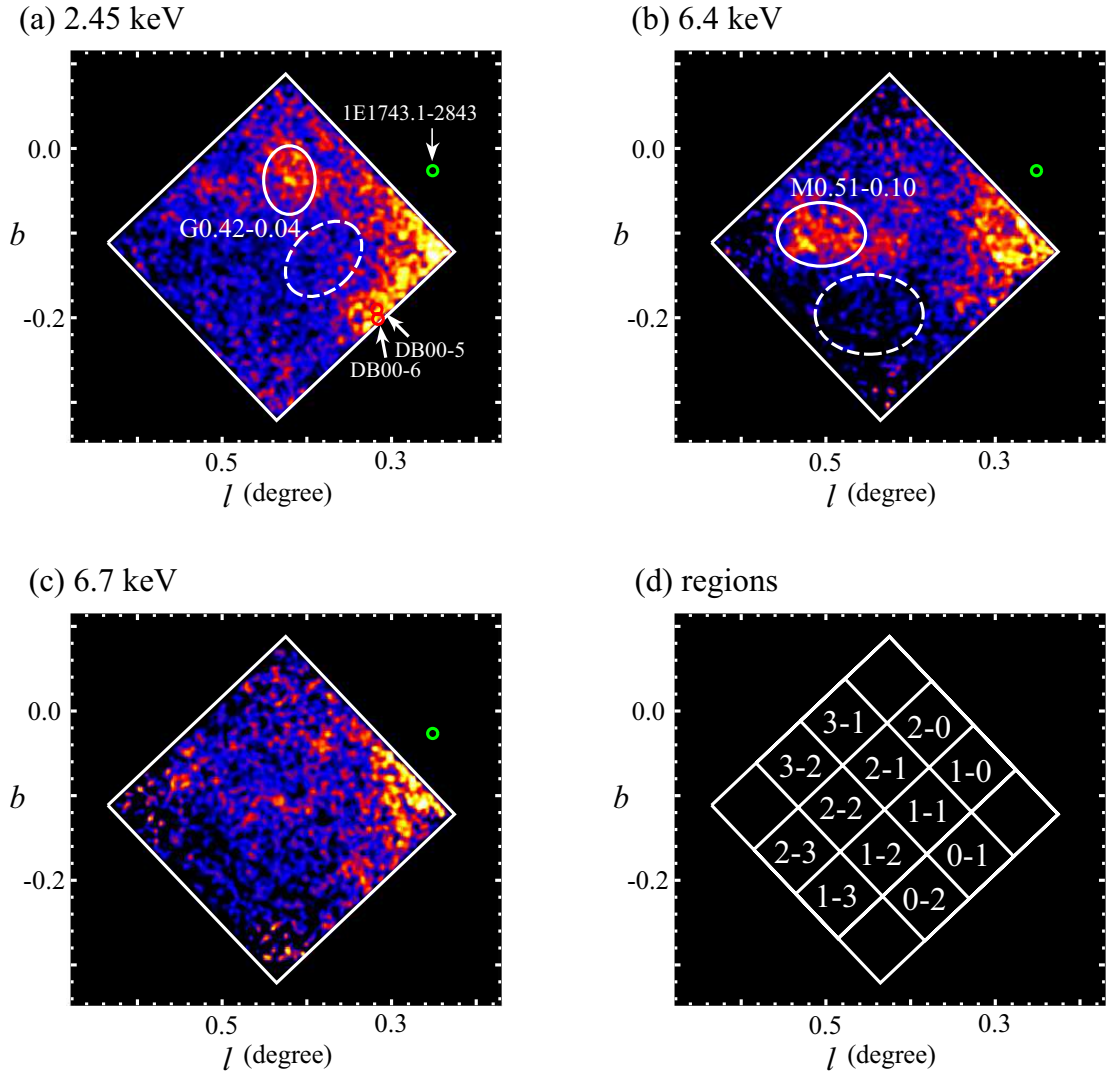}
  \end{center}
  \caption{Narrow-band images in the 2.35--2.55 keV (a),
    6.30--6.50 keV (b), and 6.57--6.77 keV (c). They are
    the co-added images of the four XIS. Subtraction of the NXBG
    had already been done, and the effects of exposure and vignetting
    were taken into account. We collected spectra of G0.42$-$0.04 and its
    background from the solid and dashed line regions in panel
    (a), respectively. The solid and dashed lines in panel (b)
    show the regions for the spectra of M0.51$-$0.10 and its background,
    respectively.
    The minor and major axes of each elliptical region are 
	$\timeform{1\prime.8}\times\timeform{2\prime.4}$ (G0.42$-$0.04: source),
	$\timeform{2\prime.2}\times\timeform{3\prime.1}$ (G0.42$-$0.04: background),
	$\timeform{2\prime.3}\times\timeform{3\prime.1}$ (M0.51$-$0.10: source),
	$\timeform{2\prime.8}\times\timeform{3\prime.8}$ (M0.51$-$0.10: background), 
    respectively.
    (d) shows the small square regions and their ID numbers used in 
    sec~\ref{sec:ironline}
	}
  \label{fig:3maps}
\end{figure*}
We examined the 
characteristic lines of Fe\emissiontype{I}, Fe\emissiontype{XXV},
Fe\emissiontype{XXVI}, and S\emissiontype{XV}. We show narrow-band 
images of 2.45~keV (2.35--2.55 keV;
S\emissiontype{XV}), 6.4~keV (6.30--6.50 keV; Fe\emissiontype{I}),
and 6.7~keV (6.57--6.77 keV; Fe\emissiontype{XXV}) in figure
\ref{fig:3maps}a, \ref{fig:3maps}b and \ref{fig:3maps}c, respectively. 
They are co-added 
images of the four XIS, in which the NXBG was subtracted and 
the correction of exposure and vignetting effects had already been
done. The exposure maps were made with {\tt xissim} \citep{Ishisaki2007arfgen} 
in HEADAS software 6.1.2.

We can see two clear diffuse sources in the two elliptic solid 
regions shown in figures~\ref{fig:3maps}a and \ref{fig:3maps}b. We hereafter call
the source in figure~\ref{fig:3maps}(a) ``Suzaku J1746.4$-$2835.4 (G0.42$-$0.04)'' 
and that in figure~\ref{fig:3maps}(b) ``Suzaku J1747.1$-$2833.2 (M0.51$-$0.10)'', 
respectively. 
G0.42$-$0.04 is a newly discovered source in this observation.
M0.51$-$0.10 is identified with the 6.4~keV cloud ``Sgr~B1'', discovered by 
\citet{Yusef2007}, whose position of 
$(l,\ b)\sim(\timeform{0\circ.5},\ \timeform{-0\circ.1})$ is the same.

We here note excesses near the western edge of the FOV in 
each image. A known bright X-ray source with a flux of
a few$\ 10^{-10}$ ergs s$^{-1}$ cm$^{-2}$, 1E 1743.1$-$2843, is located
outside the FOV in the northwest direction \citep{DelSanto20061E1743}, 
and its position is marked with a small green circle in 
figure~\ref{fig:3maps}. The northwest excess is due to the XRT
PSF (point spread function) tail of 1E 1743.1$-$2843. It
has a hard X-ray spectrum, which is consistent with the result that the 
northwest excess is seen in all panels in figure~\ref{fig:3maps}.

The southwest excess is seen only in figure \ref{fig:3maps}a.
\citet{Dutra2000} reported that two stellar clusters (DB00--5 and
DB00--6) are located at the position marked in figure
\ref{fig:3maps}a. \citet{Law2004} identified them as soft X-ray 
sources with the fluxes of $10^{-14}-10^{-13}$~ergs~s$^{-1}$~cm$^{-2}$ 
in the 2 to 10 keV band, and suggested that the stellar clusters 
are located in the foreground of the GC region. 
The preliminary analyses of our observation 
show that the southeast excess in figure \ref{fig:3maps}a have
consistent flux with those reported for the stellar clusters.

%%%%%%%%%%%%%%%%%%%%%%%%%%%%%%%%%%%%%%%%%%%%%%%%%%%%%%%%%%%%%%%%%%%%%%%%%%%%
%%%%%%%%%%%%%%%%%%%%%%%%%%%%%%%%%%%%%%%%%%%%%%%%%%%%%%%%%%%%%%%%%%%%%%%%%%%%

\subsection{Spatial Distribution of the Fe\emissiontype{XXV} K$\alpha$ Emission}
\label{sec:ironline}
\begin{figure*}[htpb]
  \begin{center}
    \FigureFile(160mm,100mm){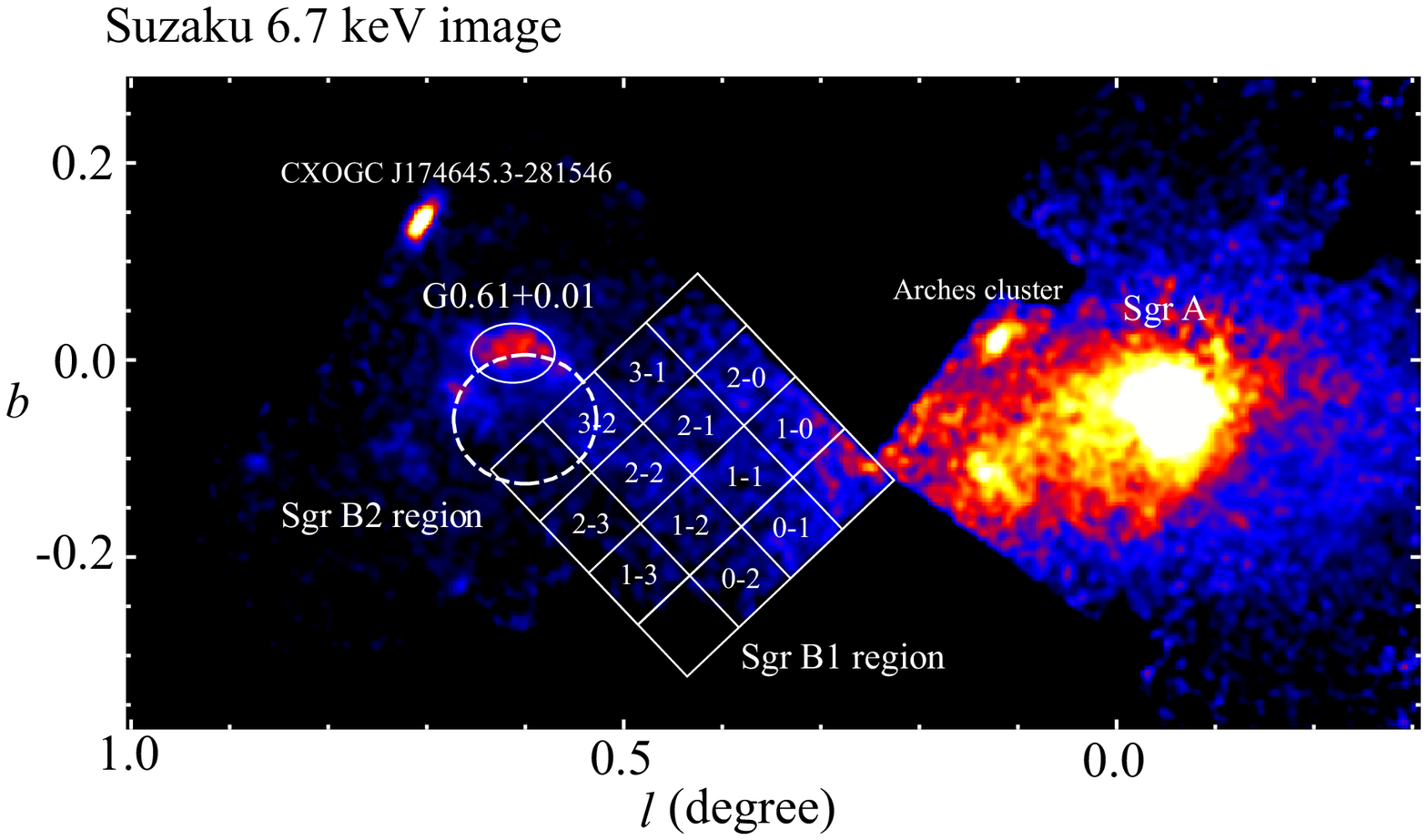}
  \end{center}
  \caption{The 6.7~keV band image of the Sgr A to Sgr B2 
    regions with Suzaku. Excepting Sgr~B1, the image is adopted from 
    \citet{Koyama2007sgrB2}, \citet{Koyama2007GCplasma}, 
    \citet{Hyodo2007GCps} and \citet{Mori2007}.
    The small square regions and the numbers are the same as in 
    figure~\ref{fig:3maps}d. G0.61+0.01 is located to the east of 
    the Sgr B1 region (the solid ellipse; \cite{Koyama2007sgrB2}). 
    We can think that G0.61+0.01 is a part of the dashed shell. 
    There are other bright sources: CXOGC J174645.3--281546
    \citep{Muno2006, Hyodo2007GCps} and 
    Arches cluster \citep{Tsujimoto2007}.
    }
  \label{fig:GC-sgrB2}
\end{figure*}

\begin{figure}[htpb]
  \begin{center}
    \FigureFile(80mm,50mm){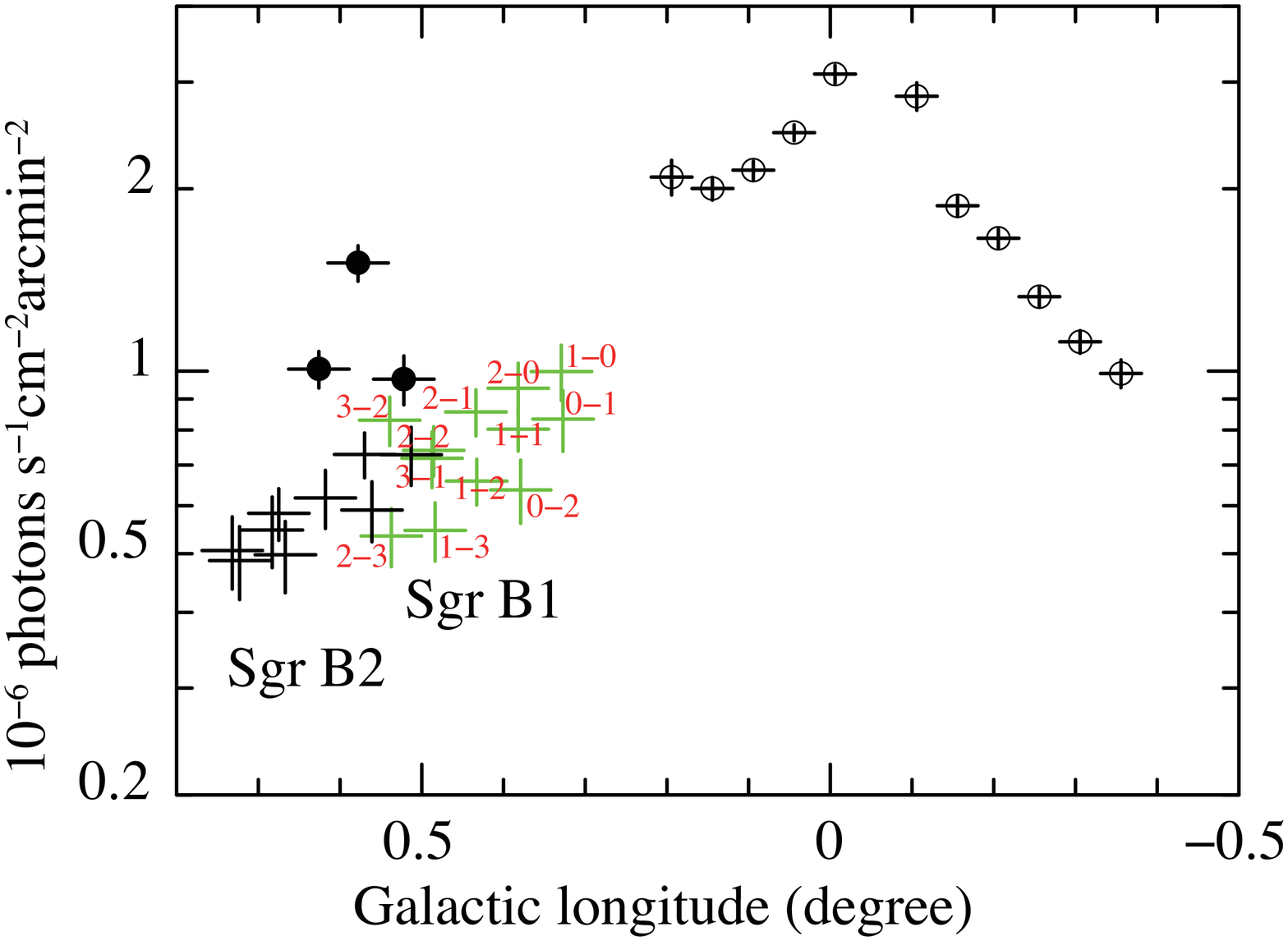}
    \FigureFile(80mm,50mm){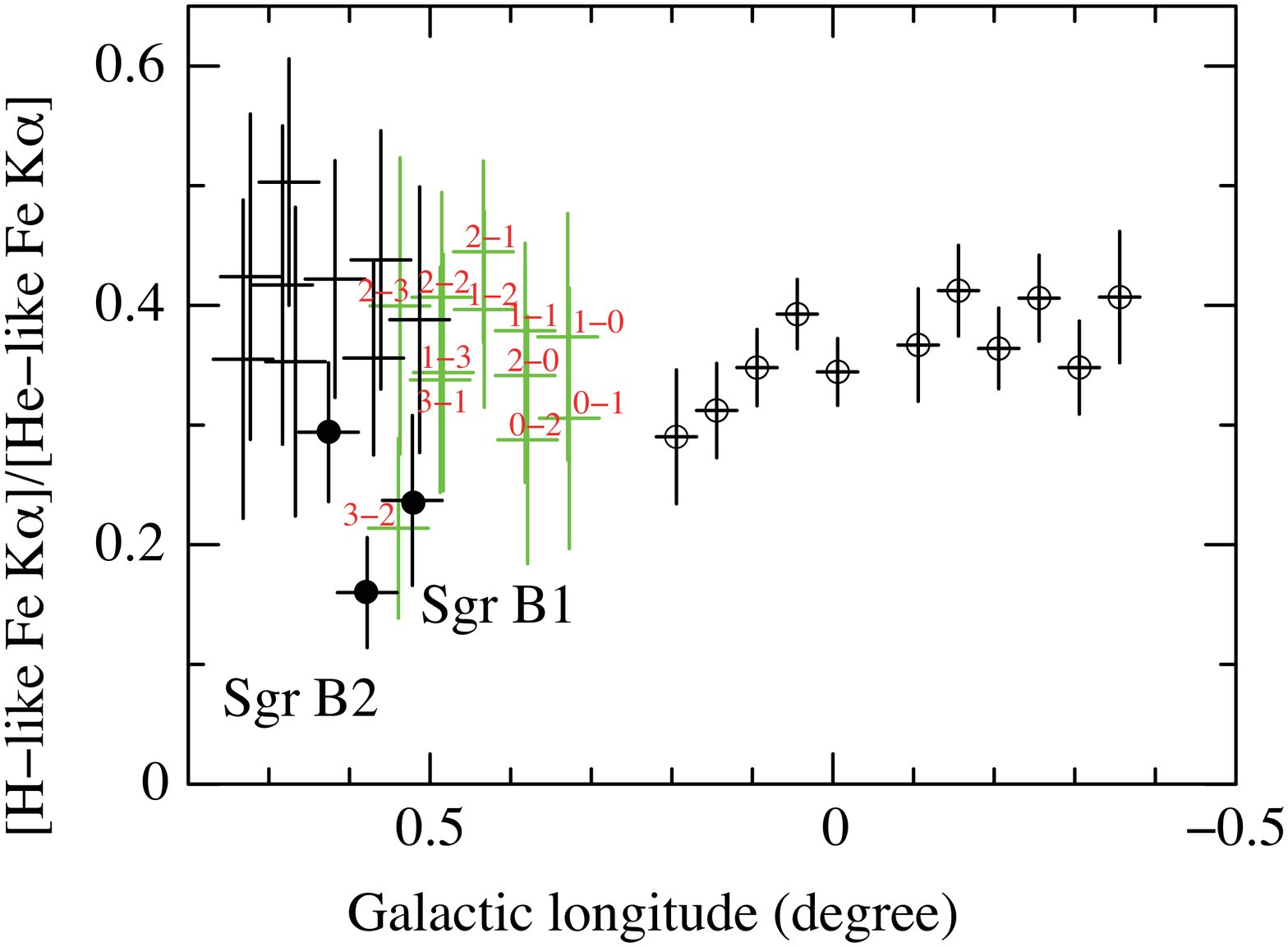}
  \end{center}
  \caption{Top panel: The 6.7 keV line fluxes of the GC region at
    $l=\timeform{-0\circ.4}-\timeform{+0\circ.8}$, including the
    Sgr~B1 and B2 regions. The data of the Sgr~B1 region obtained in
    this observation are colored with green. Data marked with open
    circles are adopted from \citet{Koyama2007GCplasma}. The data with
    filled circles correspond to SNR G0.61+0.01
    \citep{Koyama2007sgrB2}. The ID number of the data correspond 
    to the regions with the same ID number as in figure~\ref{fig:3maps}d.
    Bottom panel: The same as in the top panel, but for
    the photon flux ratios between the 6.7 and 6.9~keV lines.}
  \label{fig:connectionGC-sgrB1}
\end{figure}

Figure \ref{fig:3maps}c and figure~\ref{fig:GC-sgrB2} show a narrow-band 
image of 6.7~keV (Fe\emissiontype{XXV}), namely the distribution of the 
GCDX. Figures 4, 5, and 6 of \citet{Koyama2007GCplasma} give profiles of
the fluxes of the 6.7~keV line and the flux ratios of 6.7 
and 6.9 (Fe\emissiontype{XXVI}) keV lines as a
function of the galactic longitude between
$l=\timeform{-0\circ.4}-\timeform{+0\circ.2}$. We made the same
profiles for the Sgr~B1 region of
$l=\timeform{+0\circ.3}-\timeform{+0\circ.6}$, and compared them
with the results of \citet{Koyama2007GCplasma}.

First, we divided the Sgr~B1 region into 4 $\times$ 4 small regions,
excluding the four corners illuminated by the built-in calibration
sources, as shown in figure~\ref{fig:3maps}d. 
Next, we obtained the co-added spectrum of the 3 FIs (XIS0, 2, and 3) 
for each small region. We finally obtained 6.7~keV and 6.9~keV line 
fluxes and their
ratio for each small region by fitting the spectrum with a continuum
(thermal bremsstrahlung), Fe lines, which are 
Fe\emissiontype{I} K$\alpha$ and K$\beta$ (6.4 and 7.05 keV), 
Fe\emissiontype{XXV} K$\alpha$ (6.7 keV), and 
Fe\emissiontype{XXVI} Ly$\alpha$ (6.9 keV),
and the cosmic X-ray background (CXB). \citet{2006JPhCS..54...95K}
suggested that the 6.7~keV emission smoothly extends to the Sgr~B2 region
($l=\timeform{+0\circ.6}-\timeform{+0\circ.8}$) with almost the
same temperature. Thus, we made follow-up analyses on the Fe emission
lines for the Sgr~B2 region by the same method as was used for Sgr~B1. 

In figure~\ref{fig:connectionGC-sgrB1}, we show the 6.7~keV line flux 
(top) and the ratio between the photon fluxes of the 6.7~keV and 6.9~keV 
lines (bottom) as a function of the Galactic longitude. Excluding the 
data of the region at G0.61+0.01 (will be mentioned later on), we found 
that the 6.7~keV line flux exponentially 
decreases from the peak at the GC ($l\sim\timeform{-0\circ.05}$), and
smoothly connects the Sgr~B1 and B2 regions. The line-flux ratios of
[Fe\emissiontype{XXVI}]/[Fe\emissiontype{XXV}] in the Sgr~B1 and B2
regions are almost the same as those in
$l=\timeform{-0\circ.4}-\timeform{+0\circ.2}$. 

We note that the three data points indicated with filled circles in the
Sgr~B2 region, having high 6.7~keV fluxes and low flux ratios between
the 6.7~keV and 6.9~keV lines, correspond to SNR G0.61+0.01, which
\citet{Koyama2007sgrB2} recently discovered with Suzaku. The plasma
temperature of G0.61+0.01 is $kT\sim3$ keV \citep{Koyama2007sgrB2},
which is significantly lower than that of the GCDX ($kT\sim$6--7 keV:
\cite{Koyama2007GCplasma}). 
We can see that the 6.7 keV intensity of the small region [3--2] is rather
higher than expected from the whole tendency, and that the ratio of
[Fe\emissiontype{XXVI}]/[Fe\emissiontype{XXV}] is lower than the
average of the GCDX. G0.61+0.01 is located to the east of the
small region [3--2] (see figure~\ref{fig:GC-sgrB2}).
This result suggests that the SNR would have a
larger size than the one reported by \citet{Koyama2007sgrB2}, and that 
the small region [3--2] is a part of G0.61+0.01.

%%%%%%%%%%%%%%%%%%%%%%%%%%%%%%%%%%%%%%%%%%%%%%%%%%%%%%%%%%%%%%%%%%%%%%%%%%%%%

\subsection{Discovery of a 2.45 keV Clump, G0.42$-$0.04}
\begin{figure}
 \begin{center}
  \FigureFile(80mm,50mm){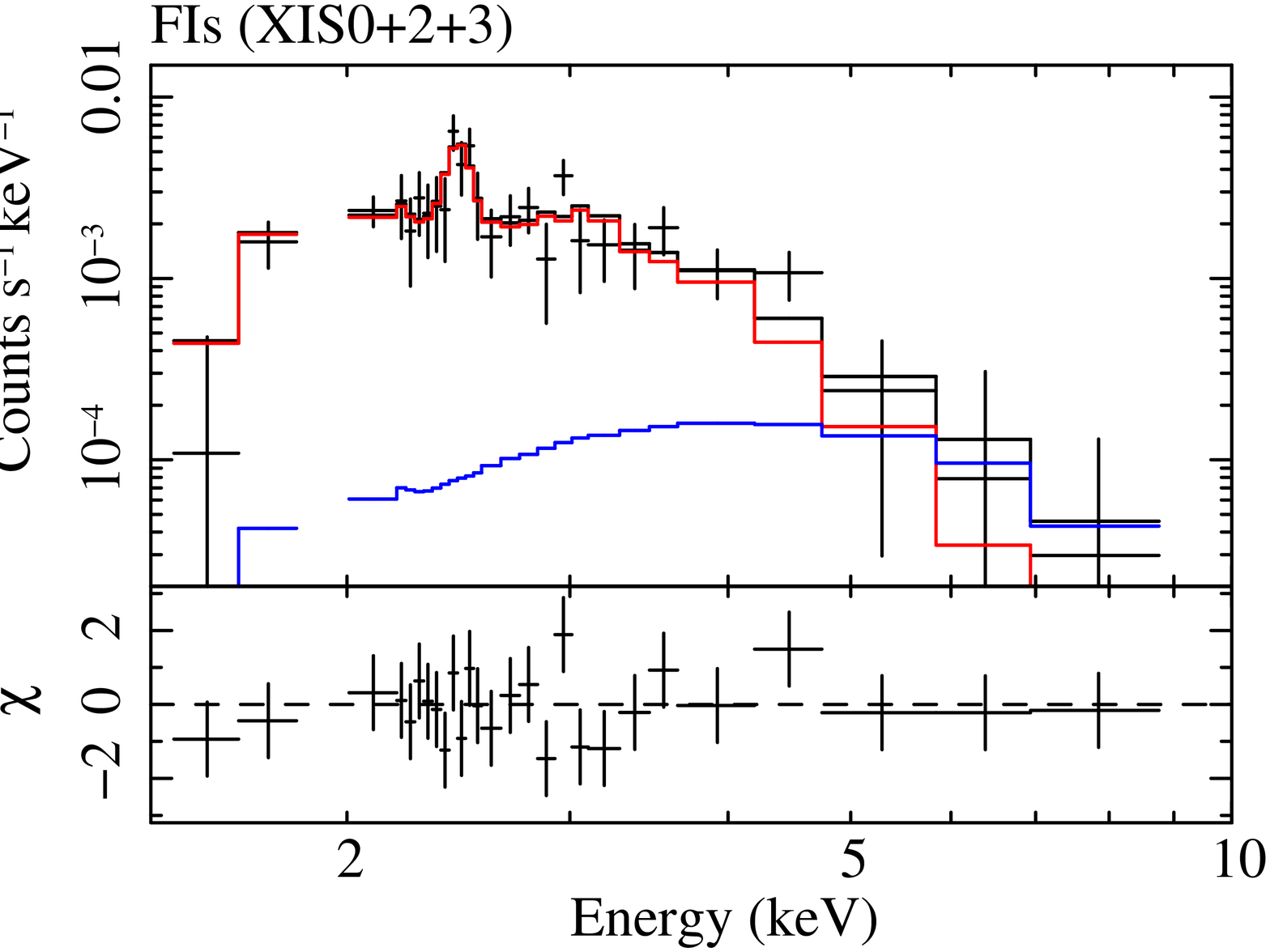}
  \FigureFile(80mm,50mm){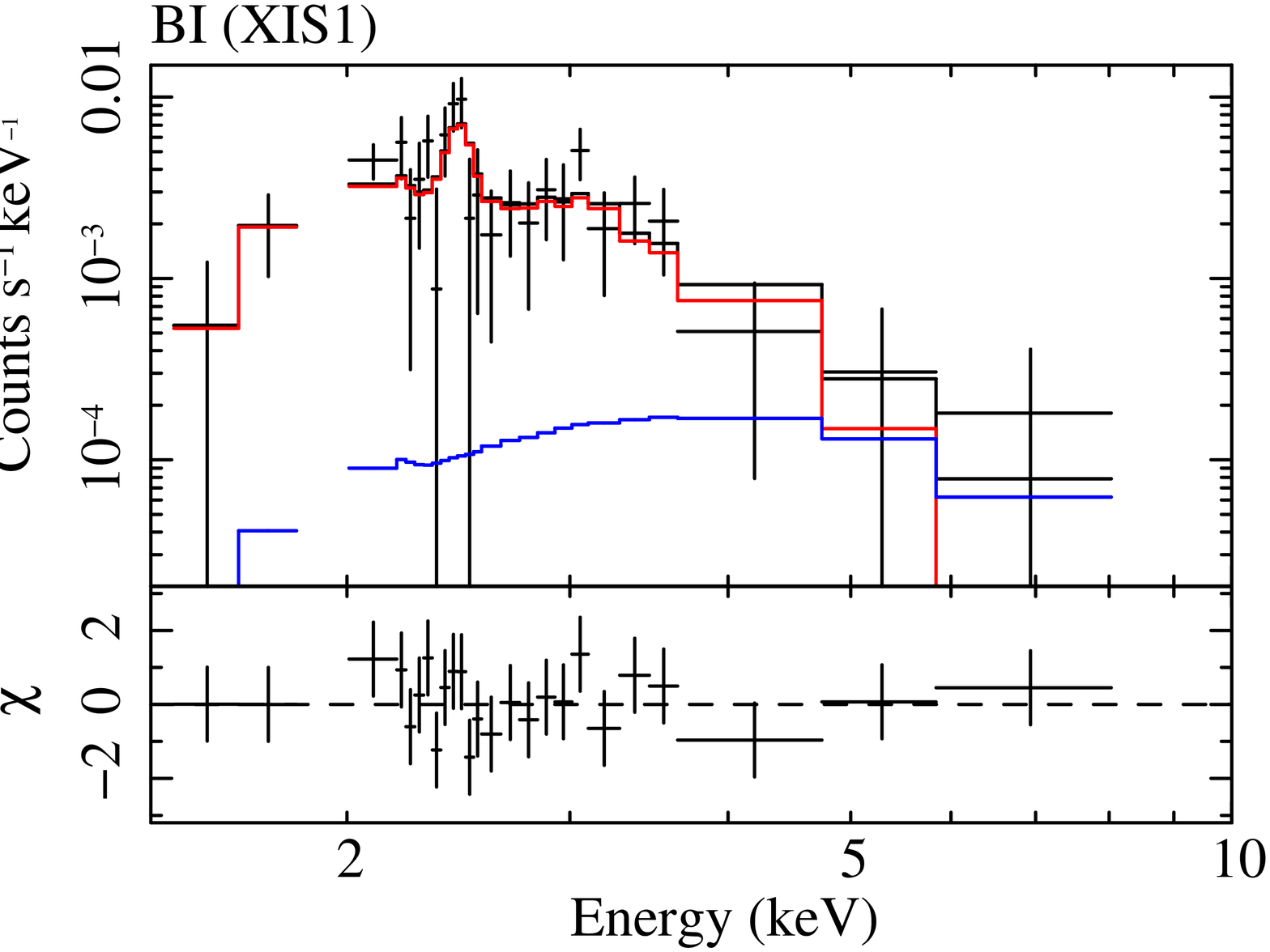}
 \end{center}
\caption{Background-subtracted FIs (top) and BI (bottom) spectra of G0.42$-$0.04. 
  The source and background data were extracted 
  from the solid and dashed regions in figure \ref{fig:3maps}a, respectively.
  The red and blue model lines show the plasma component 
  and the residual of the point sources, respectively.}
\label{fig:G0.4spectra}
\end{figure}

The background-subtracted spectra of the new source, G0.42$-$0.04, are shown
in figure~\ref{fig:G0.4spectra}. 
We ignored the energy range at the Si absorption edge due to 
the CCDs used in the XIS, where the systematic uncertainties still 
remained in the response matrices available at the time of writing this paper. 
The background for G0.42$-$0.04 consists of the NXBG, CXB, and GCDX.
As shown in the previous section, the X-ray flux from the GCDX is rather
different from position to position. We thus selected the dashed
region in figure \ref{fig:3maps}a for the background region, so that
the average line flux of the 6.7~keV line (the reference of the GCDX)
is almost the same between the region for the G0.42$-$0.04 (the
small region [2--1]) and the one for the background (the small regions
[1--0] and [1--1]).
Figure \ref{fig:G0.4spectra}
shows no residuals at the iron emission lines of 6.7 and 6.9~keV.
It is thus confirmed that background emission from the GCDX 
was successfully subtracted. 
Additionally, the NXBG and CXB are also thought to be subtracted.

Since there exist many X-ray point sources in the GC region, we next 
checked the contribution from point sources. According to the X-ray 
point-source catalog with Chandra \citep{Muno2006}, the fluxes
per square arcminute (2--8 keV)
from point sources are $9.9 \times 10^{-15}\ \ergcms {\ \rm
  arcmin}^{-2}$ (8 point sources) in the region of G0.42$-$0.04, and $7.8
\times 10^{-15}\ \ergcms {\ \rm arcmin}^{-2}$ (10 sources) in the
background region, respectively. 
The difference of $2.1 \times 10^{-15}\ \ergcms {\ \rm arcmin}^{-2}$ 
corresponds to $20$\% of the surface brightness of 
G0.42$-$0.04, itself ($1.2 \times 10^{-14}\ \ergcms {\ \rm arcmin}^{-2}$), 
observed by Suzaku.

Pronounced features of the spectra are the S\emissiontype{XV} line
(2.45 keV), a cut-off below 2 keV, and a steep slope above 4 keV. We
fitted the spectra with a simple model of an absorbed thin thermal
plasma ({\tt VAPEC} model in {\tt XSPEC}) for G0.42$-$0.04 
plus a model of an absorbed power-law component representing 
the residual spectrum of the point sources. 
We fixed the model parameters for the power-law component to be those
given in table~\ref{tab:G0.4fit} while referring to \citet{Muno2006}. 
The spectra were nicely fitted with a model with a plasma temperature of 
$\sim 0.7$ keV. The absorption column of $\sim8\times 10^{22}$ cm$^{-2}$ 
is consistent with a source in the GC region ($6 \times 10^{22}$ cm$^{-2}$), 
which suggests that G0.42$-$0.04 is located in the GC region. 
Assuming a distance of 8.5 kpc and the shape of G0.42$-$0.04 to be like a 
rugby ball, its 3-axis radii are $6$ pc $\times$ $4.5$ pc $\times$ $4.5$ pc. 

\begin{table}[t]
  \caption{Result of a spectral fitting of G0.42$-$0.04 with {\tt VAPEC} in 
  {\tt XSPEC}.}
  \begin{center}
  \label{tab:G0.4fit}
  \begin{tabular}{lc}
    \hline
    Model component& Value \\ \hline
    \ Absorption1 ($N_{\rm H}$) ($10^{22}$\ cm$^{-2}$) & $7.9^{+1.1}_{-1.5}$  \\
    \ CIE Plasma (VAPEC): &  \\
    \ \ \ Temperature $kT$ (keV) & $0.70^{+0.23}_{-0.21}$ \\
    \ \ \ Abundance Sulfur (solar) & $0.9^{+0.4}_{-0.3}$ \\ 
    \ \ \ Normalization\footnotemark[$*$] & $8.8^{+0.8}_{-0.9}$ \\
    (Residual point source component)\footnotemark[$\S$] & \\
    \ Absorption2 ($N_{\rm H}$) ($10^{22}$\ cm$^{-2}$) & $6.0$ \\
    \ Power-law: & \\
    \ \ \ Photon index $\Gamma$ & $1.5$ \\
    \ \ \ Normalization\footnotemark[$\|$] & $1.0$ \\ \hline
    Observed Flux (10$^{-13}$ ergs cm$^{-2}$ s$^{-1}$)\footnotemark[$\dagger$] & 
    $1.4^{+0.2}_{-0.1}$ \\
    Luminosity (10$^{33}$ ergs s$^{-1}$)\footnotemark[$\ddagger$] & $6.5^{+0.5}_{-0.6}$ 
    \\ \hline
    $\chi ^2$/dof & 33/48 \\
    \hline
   \multicolumn{2}{@{}l@{}}{\hbox to 0pt{\parbox{85mm}{\footnotesize
   \ 
       \par\noindent
       The uncertainties indicate the 90\% confidence levels. 
       \par\noindent
       \footnotemark[$*$] $10^{-11}/(4\pi D^2)\int n_{\rm e} n_{\rm H} dV$, 
       where $D$ is the distance to the source (cm), $n_{\rm e}$ and $n_{\rm H}$ 
       are the electron and hydrogen density (cm$^{-3}$), respectively.
       \par\noindent
      \footnotemark[$\S$] The shape of the spectrum is referred to 
       \citet{Muno2006}. These parameters are fixed. 
       \par\noindent
       \footnotemark[$\|$] The unit is $10^{-5}$ photons cm$^{-2}$ s$^{-1}$ 
       keV$^{-1}$ at 1 keV.
       \par\noindent
       \footnotemark[$\dagger$] In the 2 to 8 keV band, 
       except the residual point source component.
       \par\noindent
       \footnotemark[$\ddagger$] Absorption corrected in the 2 to 8 keV band, 
       except the residual point source component. 
       The distance toward G0.42$-$0.04 is assumed to be $8.5$ kpc.
       \par\noindent
     }}}
  \end{tabular}
  \end{center}
\end{table}

%%%%%%%%%%%%%%%%%%%%%%%%%%%%%%%%%%%%%%%%%%%%%%%%%%%%%%%%%%%%%%%%%%%%%%%%%%%%
%%%%%%%%%%%%%%%%%%%%%%%%%%%%%%%%%%%%%%%%%%%%%%%%%%%%%%%%%%%%%%%%%%%%%%%%%%%%

\subsection{Spectrum of a 6.4 keV Clump, M0.51$-$0.10}
\begin{figure}[t]
 \begin{center}
  \FigureFile(80mm,50mm){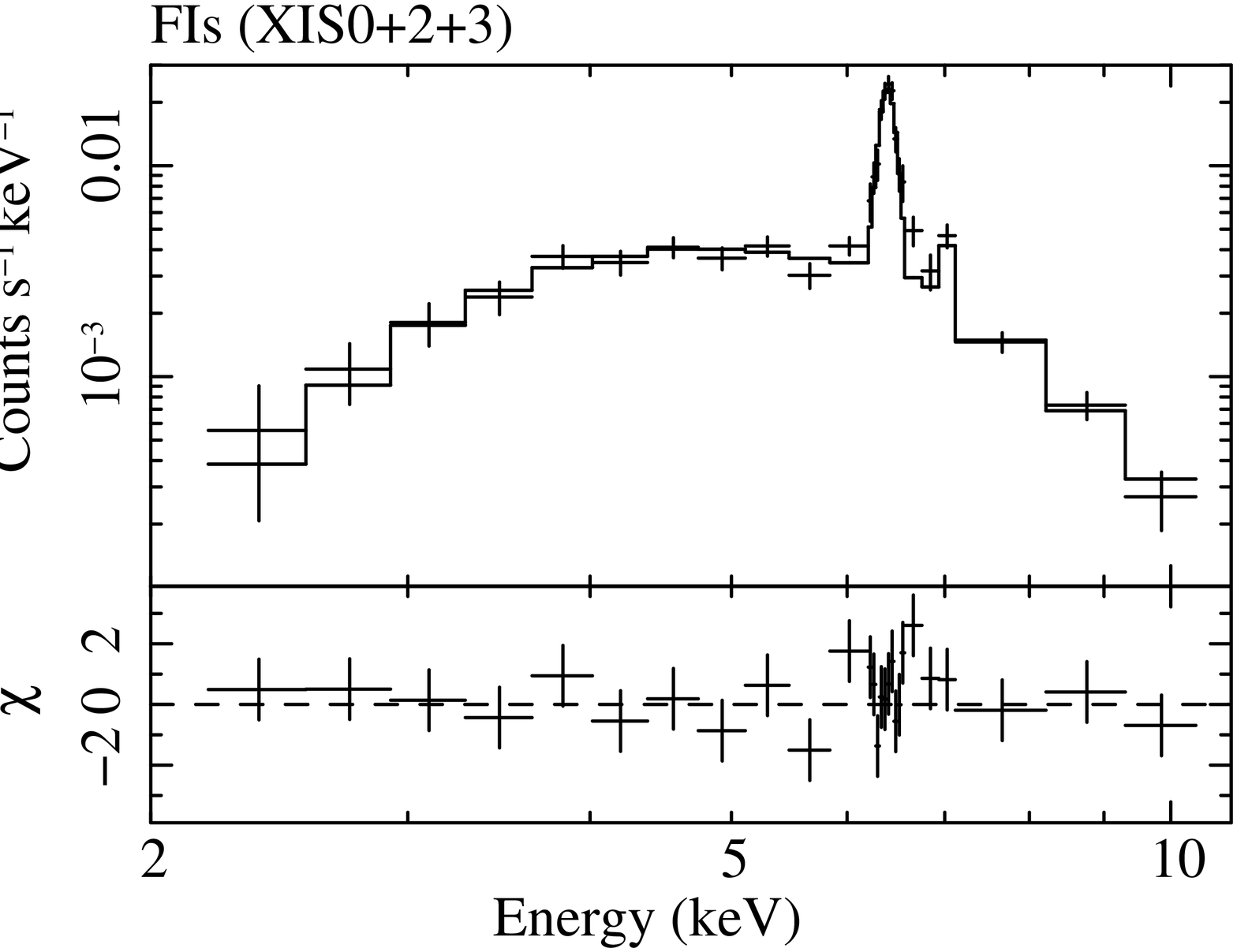}
  \FigureFile(80mm,50mm){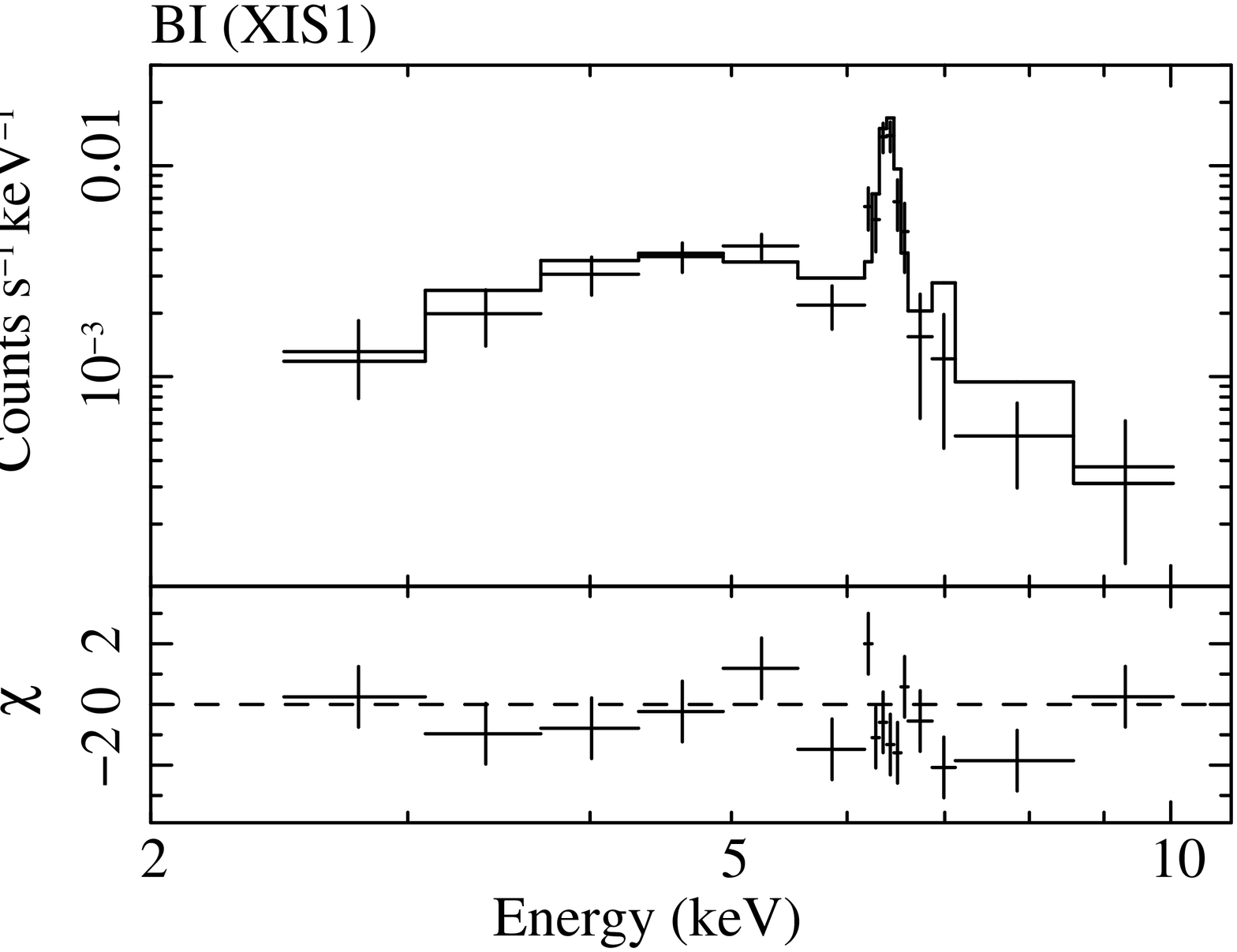}
 \end{center}
\caption{ 
The M0.51$-$0.10 spectra of the FIs (top) and BI (bottom). 
  The source and background spectra were extracted 
  from the solid and dashed region in figure \ref{fig:3maps}(b), respectively.
}
\label{fig:3-3-2}
\end{figure}
We examined M0.51$-$0.10, a diffuse source in the
6.4 keV narrow-band image of figure \ref{fig:3maps}b. 
We identified this diffuse source with the 6.4 keV cloud ``Sgr~B1'' 
detected by \citet{Yusef2007}. However, we call the 6.4~keV cloud
``M0.51$-$0.10'' in order to avoid any confusion with
the H\emissiontype{II} region ``Sgr~B1''
at $(l,\ b)=(\timeform{0\circ.5},\ \timeform{-0\circ.05})$ near the 6.4~keV cloud.

We obtained the
background-subtracted spectra shown in figure \ref{fig:3-3-2}, in 
which the source and the background-spectra were extracted from the solid
and dashed regions in figure~\ref{fig:3maps}b, respectively. 
Because we selected the background region while
referring to the top panel of figure \ref{fig:connectionGC-sgrB1},
the 6.7 keV line fluxes, namely the GCDX, of the
two regions are almost the same level within the statistical errors.
We assume that since the NXBG and CXB are the same in the source 
and background regions, their contributions were successfully subtracted.

We checked a residual contribution of point sources after the
background subtraction with the point-source catalog by
\citet{Muno2006}. There are 14 and 11 point sources in the source and
background regions, respectively. Those fluxes per square arcminute
in the 2 to 10 keV band are $5.0 \times 10^{-15}$ $\ergcms$
arcmin$^{-2}$ in the source region, and $6.4 \times 10^{-15}$
$\ergcms$ arcmin$^{-2}$ in the background region. Though the flux in
the background region is somehow larger than that in the source
region, the difference ($1.4 \times 10^{-15}$ $\ergcms$ arcmin$^{-2}$)
is only 3\% of the flux of M0.51$-$0.10 ($5.4 \times 10^{-14}$ $\ergcms$
arcmin$^{-2}$). Thus, the contribution from the point sources after
background subtraction can be ignored. 

The XIS spectra of M0.51$-$0.10 contain a prominent 6.4 keV
emission line. We fitted the spectra with a model consisting of 
an absorbed power-law component plus two Gaussian lines of 
Fe\emissiontype{I} K$\alpha$ (6.4 keV) and K$\beta$ (7.06 keV). 
We obtained an acceptable result, 
and give the best-fit parameters in table~\ref{tab:sgrB1}.
\citet{Yusef2007} reported that the equivalent width of the Fe\emissiontype{I} 
K$\alpha$ line is $\sim0.6$ keV, smaller than our results of $\sim1.4$ keV.
\citet{Yusef2007} subtracted only the NXBG from the source spectrum.
We examined that the equivalent width of the Fe\emissiontype{I} K$\alpha$ line
was obtained to be $\sim0.5$ keV from our spectra when we subtracted 
only the NXBG from the source spectra.
We believe that the difference between the result of \citet{Yusef2007} 
and ours is due to different assumptions concerning the background emission.

\begin{table}[t]
  \caption{Result of a spectral fitting of the M0.51$-$0.10. The model is an 
  absorbed power-law plus two Gaussian lines.}
  \begin{center}
    \label{tab:sgrB1}
    \begin{tabular}{lc}
      \hline
      Model component& Value \\ \hline
      Absorption ($N_{\rm H}$) ($10^{23}$\ cm$^{-2}$) & $1.5^{+0.2}_{-0.1}$  \\
      Continuum (power-law): &  \\
      \ \ Photon index ($\Gamma$) & $1.8^{+0.4}_{-0.5}$ \\
      Gaussian1 (Fe\emissiontype{I} K$\alpha$): & \\
      \ \ Line center energy (eV) & $6402^{+6}_{-7}$ \\
      \ \ Intensity (10$^{-5}$ photons cm$^{-2}$ s$^{-1}$) & $2.8^{+0.2}_{-0.4}$ \\
      \ \ Equivalent Width (keV) & $1.4^{+0.3}_{-0.3}$ \\
      Gaussian2 (Fe\emissiontype{I} K$\beta$): & \\
      \ \ Line center energy (eV)\footnotemark[$*$] & $7061$ \\
      \ \ Intensity (10$^{-5}$ photons cm$^{-2}$ s$^{-1}$) & $0.44^{+0.17}_{-0.21}$ \\
      \ \ Equivalent Width (keV) & $0.23^{+0.22}_{-0.21}$ \\ \hline
      Observed Flux (10$^{-12}$ ergs cm$^{-2}$ s$^{-1}$)\footnotemark[$\dagger$] & 
      $1.2^{+0.1}_{-0.1}$ \\
      Luminosity (10$^{34}$ ergs s$^{-1}$)\footnotemark[$\ddagger$] & $1.9^{+0.1}_{-0.1}$ 
      \\ \hline
      $\chi ^2$/dof & 50/37 \\
      \hline
   \multicolumn{2}{@{}l@{}}{\hbox to 0pt{\parbox{85mm}{\footnotesize
\
       \par\noindent
       The uncertainties indicate the 90\% confidence limits.
       \par\noindent
       \footnotemark[$*$] Fixed to $1.103\times E$(Fe\emissiontype{I} K$\alpha$).
       \par\noindent
       \footnotemark[$\dagger$] In the 2 to 10 keV band.
       \par\noindent
       \footnotemark[$\ddagger$] Absorption corrected in the 2 to 10 keV band.
     }}}
    \end{tabular}
  \end{center}
\end{table}

%%%%%%%%%%%%%%%%%%%%%%%%%%%%%%%%%%%%%%%%%%%%%%%%%%%%%%%%%%%%%%%%%%%%%%%%%%%%
%%%%%%%%%%%%%%%%%%%%%%%%%%%%%%%%%%%%%%%%%%%%%%%%%%%%%%%%%%%%%%%%%%%%%%%%%%%%

%%%%%%%%%%%%%%%%
%% Discussion %%
%%%%%%%%%%%%%%%%

\section{Discussion}

\subsection{The Thermal Plasma Clump, G0.42-0.04}
We here discuss the nature of G0.42$-$0.04 as an SNR. 
Assuming an uniform plasma filling the rugby ball region, we obtained the 
physical parameters of G0.42$-$0.04 given in table~\ref{tab:G0.4physicalvalue}.
The plasma temperature of $kT\sim 0.7$~keV and the size of $\sim10$~pc
are typical values for an SNR. On the other hand, the thermal energy of 
$4\times 10^{49}$ ergs is lower than the canonical value of 
$0.7\times E_{\rm total} \sim 0.7\times10^{51}$~ergs for 
an ordinary SNR in the Sedov phase. 
The dynamical timescale can be obtained to be $\sim 8000$ years from 
the sound velocity of $7\times 10^7$ cm s$^{-1}$ for a plasma with 
$kT\sim 0.7$~keV and the size of G0.42$-$0.04. Assuming the timescale 
as the age of an SNR, the size of $\sim10$ pc and the total 
mass of $12~\MO$ for G0.42$-$0.04 are smaller than the values of 
$\sim20$ pc and $\sim100~\MO$ for a typical SNR having a similar age 
in the Sedov phase. A general structure of an SNR is that the temperature of
the center is higher than that in the shell, while the surface 
brightness in the center is much lower than that in the shell. 
Thus, it is possible that the soft X-ray emission from the shell 
of G0.42$-$0.04 would be absorbed by the large absorption column 
toward the GC. Thus, the results of our observation are consistent 
with an SNR. We conclude that G0.42$-$0.04 is an SNR candidate. 

\begin{table}[t]
  \caption{Physical parameters of G0.42$-$0.04.}
  \begin{center}
  \label{tab:G0.4physicalvalue}
  \begin{tabular}{lc}
    \hline
    Parameter\hspace{30mm}\ & Value \\ \hline
    EM\footnotemark[$*$] (cm$^{-3}$) & $8 \times 10^{57}$  \\
    $n_{\rm e}$\footnotemark[$\dagger$] (cm$^{-3}$) & 0.8 \\
    $M$\footnotemark[$\ddagger$] (\Mo)& 12 \\
    $E_{\rm thermal}$\footnotemark[$\S$] (ergs) & $4\times10^{49}$ \\
    $t_{\rm dyn}$\footnotemark[$\|$] (year) & $8000$ \\
    \hline
   \multicolumn{2}{@{}l@{}}{\hbox to 0pt{\parbox{70mm}{\footnotesize
\ 
       \par\noindent
       The plasma is assumed to be uniform density and elliptical 
       shape with 3-axis radii of $6$ pc $\times$ $4.5$ pc $\times$ $4.5$ pc.
       \par\noindent
       \footnotemark[$*$] The emission measure (EM) 
       $=n_{\rm e}n_{\rm H}V$, where $n_{\rm e}$ and 
       $n_{\rm H}$ are the electron and hydrogen densities 
       and assumed to be equal. 
       $V$ is the plasma volume.
       \par\noindent
       \footnotemark[$\dagger$] The electron density.
       \par\noindent
       \footnotemark[$\ddagger$] The total mass $=n_{\rm e}m_{\rm p}V$, where 
       $m_{\rm p}$ is the proton mass.
       \par\noindent
       \footnotemark[$\S$] The thermal energy $E_{\rm thermal}=3n_{\rm e}kTV$.
       \par\noindent
       \footnotemark[$\|$] The dynamical time scale. The sound velocity of the 
       $\sim 0.7$ keV plasma is $7\times 10^{7}$ cm s$^{-1}$.
     }}}
  \end{tabular}
  \end{center}
\end{table}

We examine another possibility, that G0.42$-$0.04 is a stellar cluster. 
The temperature of $kT\sim0.7$ keV is in the range of the typical value
of $0.5\sim3$ keV for a stellar cluster. The X-ray luminosity, 
$6\times10^{33}$~ergs~s$^{-1}$, of G0.42$-$0.04 is similar to 
$2\times10^{34}$~ergs~s$^{-1}$ for the Arches cluster \citep{Tsujimoto2007} 
and $1\times10^{34}$ ergs s$^{-1}$ for the Quintuplet cluster \citep{Law2004}, 
both of which are famous stellar clusters in the GC region. 
The absorption column toward G0.42--0.04 ($\sim8\times10^{22}$ cm$^{-2}$)
is rather less than that toward the Arches cluster ($\sim14\times10^{22}$~
cm$^{-2}$; \cite{Tsujimoto2007}).
\citet{Dutra2000} searched for stellar clusters in a field of 
$\timeform{5\circ}\times\timeform{5\circ}$, centred close to the Galactic
Nucleus using the infrared 2MASS Survey archive. In their catalog, 
no cluster is reported for the position of G0.42$-$0.04.
Thus, it is unlikely that G0.42$-$0.04 is a stellar cluster. 

%%%%%%%%%%%%%%%%%%%%%%%%%%%%%%%%%%%%%%%%%%%%%%%%%%%%%%%%%%%%%%%%%%%%%
%% calclation of the required luminosity of the irradiating source %%
%%%%%%%%%%%%%%%%%%%%%%%%%%%%%%%%%%%%%%%%%%%%%%%%%%%%%%%%%%%%%%%%%%%%%

\subsection{The 6.4~keV Clump, M0.51--0.10}

The absorption column of $1.5 \times 10^{23}\ {\rm cm}^{-2}$ is 2-times
or more larger than the typical value of $6 \times 10^{22}\ {\rm cm}^{-2}$
for sources in the GC region. The strong 6.4 keV emission line implies that 
a large amount of iron in the neutral state exists at M0.51$-$0.10. Thus, 
these results suggest M0.51$-$0.10 is a local cool dense cloud in the GC
region. Indeed, a molecular cloud lies at the same position of
M0.51$-$0.10, which is a part of the Sgr~B1 molecular shell proposed by
\citet{Sofue1990}. 

Two possible scenarios for the the origin of the 6.4 keV emission line have 
been proposed. One is photo-ionization by X-rays, namely the XRN
scenario. The other is inner-shell ionization by the collision of low-energy 
cosmic-ray electrons. The observed large equivalent width of $1.4$ keV and
an absorption column reaching $\sim 10^{23}$ ${\rm cm}^{-2}$ are consistent 
with those of the XRN scenario. On the other hand, the scenario of electron 
collision expects a rather small equivalent width of $\sim 300$ eV and an
absorption column of less than $\sim 10^{21-22}$ cm$^{-2}$, assuming that 
the elemental composition is the same as that of the solar system
\citep{Tatischeff2002}. The electron-collision scenario requires an iron 
over-abundance by a factor of 4 -- 5, while the photo-ionization model does 
not require an iron over-abundance. Thus, we conclude that M0.51$-$0.10 is 
likely to be an X-ray reflection nebula.

%%%%%%%%%%%%%%%%%%%%%%%%%%%%%%%%%%%%%%%%%%%%%%%%%%%%%%%%%%%%%%%%%
%%%%%%%%%%%%%%%%%%%%%%%%%%%%%%%%%%%%%%%%%%%%%%%%%%%%%%%%%%%%%%%%%

\subsection{X-Ray Source Irradiating M0.51$-$0.10}
In order to examine the irradiating source, we estimate its luminosity
to require the M0.51$-$0.10 fluorescence. The XRN absorbs X-rays with 
energies higher than 7.1 keV from the external source, and emits a 
fluorescent line of 6.4 keV X-rays. The photon flux of the 6.4 keV
line, $C_{6.4{\rm keV}}$~(photons s$^{-1}$ cm$^{-2}$), is described by 
the following equation: 
\begin{eqnarray}
  C_{6.4{\rm keV}} &=&
  \epsilon \frac{\Omega}{4\pi} \int^{\infty}_{7.1{\rm keV}}
  \left( 1-e^{-N_{\rm Fe}\sigma_{{\rm Fe(}E{\rm )}}} \right)
  AE^{-\Gamma}dE.
  \label{eq:6.4keVemission}
\end{eqnarray}
Here, we assume that the irradiating source has a power-law spectrum
($AE^{-\Gamma}$) and the iron abundance of the molecular cloud of
M0.51$-$0.10 is solar ($[\rm Fe]$/$[\rm H]=3\times10^{-5}$). Since the
typical absorption column for sources in the GC region is $N_{\rm H} = 0.6\times
10^{23}\ {\rm cm}^{-2}$, that of M0.51$-$0.10, itself, is estimated to be
$N_{\rm H} = 0.9\times 10^{23}\ {\rm cm}^{-2}$ from the results of the
spectral fitting. $\Omega=\pi (r/D)^2$ is the solid angle covered by M0.51$-$0.10
from the view point of the irradiating source, where $r$ is
the radius of M0.51$-$0.10 and $D$ is the distance between M0.51$-$0.10
and the irradiating source; $\epsilon$ is a fluorescent yield of
0.34 for iron atom. The K-shell photo-ionization cross section per an
iron atom against an X-ray with energy above $7.1$ keV is
$\sigma_{\rm Fe}=6.0\times 10^{-18}(E/{\rm 1keV})^{-2.58}\ (\rm cm^{2})$
\citep{Henke1982}.  $C_{6.4{\rm keV}}$ is the observed
Fe\emissiontype{I} K$\alpha$ line flux of $2.8 \times 10^{-5}\ {\rm 
  photons}\ {\rm cm}^{-2}\ {\rm s}^{-1}$.
\begin{table*}
  \caption{Comparison between required luminosities and observed luminosities.}
  \begin{center}
    \label{tab:reqlumin}
    \begin{tabular}{lcccc}
      \hline
      Irradiating source & Distance\footnotemark[$*$] & Photon index & 
      $L_{\rm obs}$\footnotemark[$\dagger$] & $L_{\rm req}$\footnotemark[$\S$] \\
      \ \ \ candidates       & $D$~(pc) & $\Gamma$ & (ergs s$^{-1}$) & (ergs s$^{-1}$) \\
      \hline
      1E 1743.1$-$2843 & $40$ & $1.9$\footnotemark[$\dagger$] & $3 \times 10^{36}$ & 
      $2 \times 10^{38}$ \\
      Sgr A*        & $100$ & $2.0$\footnotemark[$\ddagger$]  & $2 \times 10^{33}$ & 
      $2 \times 10^{39}$ \\
      \hline
    \multicolumn{2}{@{}l@{}}{\hbox to 0pt{\parbox{115mm}{\footnotesize
\
	\par\noindent
	Those source are assumed to be at the same distance as M0.51$-$0.10, 
	$8.5$ kpc. The bright source 1E 1743.1$-$2843 is a neutron star or black 
	hole LMXB. Sgr~A* is a supermassive black hole at the center of our galaxy. 
        \par\noindent
        \footnotemark[$*$] Distance to M0.51$-$0.10.
        \par\noindent
        \footnotemark[$\dagger$] Observed photon index and luminosity 
	\citep{Porquet2003-1E1743, Baganoff2003sgrA*}.
        \par\noindent
        \footnotemark[$\ddagger$] Assumed value by \citet{Koyama1996} and 
	\citet{Murakami2000sgrB2}.
        \par\noindent
        \footnotemark[$\S$] Required luminosity to account for the observed 
	6.4 keV flux of M0.51$-$0.10. It is obtained from equation 
	(\ref{eq:6.4keVemission}).
      }}}
    \end{tabular}
  \end{center}
\end{table*}

We discuss candidates for the irradiating source by estimating its
luminosity with equation~(\ref{eq:6.4keVemission}). One is a bright
X-ray source, 1E 1743.1$-$2843, the PSF tail of which is seen at the
northwestern edge of the XIS FOV in figure~\ref{fig:3maps}.
1E 1743.1$-$2843 is reported to be a neutron star or a black-hole 
low-mass X-ray binary in the GC region, or toward the GC region, but
beyond there \citep{Porquet2003-1E1743, DelSanto20061E1743}.  

In the case that 1E 1743.1$-$2843 is located in the GC region, the
projected distance to the M0.51$-$0.10 XRN is $D=40\ {\rm pc}$. 
Since the power-law slope for the X-ray spectrum of 
1E 1743.1$-$2843 is reported to be $\Gamma=1.9$ by \citet{Porquet2003-1E1743},
we obtained that a luminosity of $2 \times 10^{38}\ {\rm ergs}\ {\rm s}^{-1}$ 
is required for the 6.4~keV line flux of M0.51$-$0.10. 
However, 1E 1743.1$-$2843 persistently has a luminosity of 
$(1-4)\times 10^{36}$ ergs s$^{-1}$
(e.g., \cite{Cremonesi1999, Porquet2003-1E1743}), which is 2-orders
lower than the required luminosity. In addition, no bursts have been
observed from 1E 1743.1$-$2843 in extensive observations over the last 20
years. In the case that 1E 1743.1$-$2843 would be beyond the GC region, 
it is too far to account for the 6.4~keV line flux of M0.51$-$0.10.
Thus, 1E 1743.1$-$2843 is unlikely to be the irradiating source of
M0.51$-$0.10.

The total luminosity of all cataloged bright point sources within 50
pc ($20^{\prime}$) from M0.51$-$0.10 is $\sim3\times10^{35}$ ergs s$^{-1}$,
which is far lower than the required luminosity of $10^{38-39}\ {\rm
  ergs}\ {\rm s}^{-1}$. 

\citet{Koyama1996}, \citet{Murakami2000sgrB2}, and
\citet{Murakami2001ASCAsgrC} have proposed a GC supermassive black hole,
Sgr~A*, as the irradiating source of other X-ray reflection nebulae,
Sgr B2 and Sgr C. Sgr~A* is then thought to have been $\sim 10^6$ times 
brighter about 300 years ago, the light traveling time between Sgr~B2 and Sgr~A*.

We applied the XRN scenario by past activity of Sgr A* to M0.51$-$0.10.
The Sgr~B molecular complex, consisting of Sgr B1 ($l\sim\timeform{0\circ.5}$)
and B2 ($l\sim\timeform{0\circ.8}$), belongs to the ``Galactic-Center molecular 
Arm'' surrounding Sgr~A* \citep{Sofue1995}. Sgr~B1 and Sgr~B2 are
located at the same three-dimensional distance from Sgr~A*
($\sim100$ pc) though the projected distances are different. 
The photon index for Sgr A* is reported to be $\Gamma=2.7$ 
by \citet{Baganoff2003sgrA*}, which is not consistent with 
that of M0.51$-$0.10 ($\Gamma=1.3-2.2$).
However, \citet{Baganoff2001} reported that the photon index became 
hard ($\Gamma\sim1.0$) when Sgr A* was in a flare-up state.
\citet{Koyama1996} and \citet{Murakami2000sgrB2} assumed that 
the photon index is 2.0 for the power-law spectrum of Sgr A*,
 and reported that
the required luminosity for Sgr B2 XRN is $\sim 3\times 10^{39}$ ergs s$^{-1}$.
With the same assumption, we estimated the required luminosity for M0.51$-$0.10
to be $\sim 2\times 10^{39}$ ergs s$^{-1}$. 
Thus, it is consistent with the required luminosity for Sgr~B2.
Therefore, the XRN scenario by the past activity of Sgr A*, which was
successfully applied to Sgr B2, may also be applied to M0.51$-$0.10.
Taking into account that Sgr~B1 and Sgr~B2 have the same three-dimensional 
distance from Sgr~A*, we presume that both of the two XRNe
originated in the identical past activity of Sgr~A* about 300 years ago.

%%%%%%%%%%%%%%%%%%%%%%%%%%%%%%%%%%%%%%%%%%%%%%%%%%%%%%%%%%%%%%%%%%%%%%%%%%%%%%%%%%%%%%%%
%%%%%%%%%%%%%%%%%%%%%%%%%%%%%%%%%%%%%%%%%%%%%%%%%%%%%%%%%%%%%%%%%%%%%%%%%%%%%%%%%%%%%%%%

\section{Summary}

With Suzaku, we observed of the Sgr~B1 region with an exposure time 
of about 100 ksec in March 2006. Thanks to the excellent energy 
resolution of the XIS aboard Suzaku, the large effective area and a 
long exposure, we successfully revealed the distribution of the GCDX, 
and discovered a new diffuse faint source, ``G0.42$-$0.04'', and 
obtained an excellent spectrum from the 6.4~keV cloud, ``M0.51$-$0.10''.

\begin{itemize}
\item
The line-flux ratio [Fe\emissiontype{XXVI}]/[Fe\emissiontype{XXV}] of
the Sgr~B1 and B2 regions
($l=\timeform{+0\circ.3}--\timeform{+0\circ.8}$) is consistent with
that of the region at
$l=\timeform{-0\circ.4}--\timeform{+0\circ.2}$. The Fe\emissiontype{XXV} 
line flux exponentially decreases from the GC center to the Sgr~B1 and 
B2 regions. This result suggests that the GCDX extends at least up to 
the Sgr~B1 and B2 regions with a constant temperature of $kT\sim6-7$ keV.

\item
The large absorption column of $N_{\rm H}\sim8\times 10^{22}$ cm$^{-2}$
toward G0.42$-$0.04 suggests that the source is located in the GC region. 
The spectra of G0.42$-$0.04 are well fitted with a thin thermal-plasma 
model of $kT\sim0.7$ keV. In consideration of the large absorption column 
toward the GC region, the observational results are consistent with the 
idea that G0.42$-$0.04 might be a part of an SNR. On the other hand, 
G0.42$-$0.04 is unlikely to be a stellar cluster, since no infrared stellar 
cluster has been reported at the position of G0.42$-$0.04.

\item
M0.51$-$0.10 is identified with the 6.4~keV cloud ``Sgr~B1''
detected by \citet{Yusef2007}.
The X-ray spectra of M0.51$-$0.10 exhibit a absorption column of
$N_{\rm H}\sim1.5\times10^{23}$ cm$^{-2}$ and a large equivalent width of 
$\sim1.4$ keV for the Fe\emissiontype{I} K$\alpha$ fluorescence emission line. 
This suggests that the 6.4~keV X-ray emission is unlikely to be due to the 
collision of low-energy cosmic-ray electrons, but due to the reflection 
of external hard X-rays, and hence M0.51$-$0.10 is thought to be an
X-ray reflection nebula.
The bright X-ray point source 1E~1743.1$-$2843,
or the collection of fainter point sources near M0.51$-$0.10 
does not explain the luminosity of $\gtrsim10^{38-39}$ergs s$^{-1}$
required for the 6.4 keV X-rays from M0.51$-$0.10. 
The XRN scenario based on the past activity of Sgr~A*, which 
was successfully applied to Sgr B2, may also be applied to M0.51$-$0.10.

\end{itemize}

%%%%%%%%%%%%%%%%%%%%%%%%%%%%%%%%%%%%%%%%%%%%%%%%%%%%%%%%%%%%%%%%%%%%%%%%%%%
%%%%%%%%%%%%%%%%%%%%%%%%%%%%%%%%%%%%%%%%%%%%%%%%%%%%%%%%%%%%%%%%%%%%%%%%%%%

\bigskip
%%\section*{Acknowledgment}
We are grateful to all members of the Suzaku hardware and software
teams and the science working group. HN, TI and YH are supported by 
Japan Society for the Promotion of Science (JSPS) 
Research Fellowship for Young Scientists. 
This work is based by a Grant-in-Aid for the 21st
Century COE ``Center for Diversity and Universality in Physics'' from
Ministry of Education, Culture, Sports, Science and Technology (MEXT)
of Japan, and is supported by a Grant-in-Aid for Scientific Research 
on Priority Areas in Japan (Fiscal Year 2002-2006); 
``New Development in Black Hole Astronomy''. 
This work is also partially supported by Grants-in-Aid for Scientific
Research of JSPS Nos. 18204015, 18540228, 18740105. 
% Koyama A, Yamauchi C, Matsumoto Watate B

%%%%%%%%%%%%%%%%%%%%%%%%%%%%%%%%%%%%%%%%%%%%%%%%%%%%%%%%%%%%%%%%%%%%%%%%%%%%%%%%
%%%%%%%%%%%%%%%%%%%%%%%%%%%%%%%%%%%%%%%%%%%%%%%%%%%%%%%%%%%%%%%%%%%%%%%%%%%%%%%%

\end{document}